\documentstyle[12pt,psfig,epsfig]{article}

\newcommand {\be}{\begin{equation}}
\newcommand {\ee}{\end{equation}}
\newcommand {\bea}{\begin{eqnarray}}
\newcommand {\eea}{\end{eqnarray}}

\begin{document}

\begin{titlepage}

\title{Kulback-Leibler and Renormalized Entropy: Applications to 
   EEGs of Epilepsy Patients}

\vspace{2cm}

\author{R. Quian Quiroga$^{\dagger}$, J. Arnhold$^{\dagger\ddagger}$,
  K. Lehnertz$^\ddagger$, \\ and P. Grassberger$^\dagger$\\
$^\dagger${\sl John von Neumann Institute for Computing,}\\
{\sl Forschungszentrum J\"ulich GmbH,}\\
{\sl D - 52425 J\"ulich, Germany}\\
$^\ddagger${\sl Clinic of Epileptology, University of Bonn,}\\{\sl Sigmund-Freud Str. 25,}\\
{\sl D - 53105 Bonn, Germany} }
\maketitle

\newpage




\begin{abstract}
Recently, ``renormalized entropy" was proposed as a novel measure of 
relative entropy (P. Saparin {\it et al.}, Chaos, Solitons \& Fractals 
{\bf 4}, 1907 (1994)) and applied to several 
physiological time sequences, including EEGs of patients with epilepsy. 
We show here that this measure is just a modified Kullback-Leibler (K-L)
relative entropy, and it gives similar numerical results to the standard
K-L entropy. The latter 
better distinguishes frequency contents of e.g. 
seizure and background EEGs than renormalized entropy.
We thus propose that renormalized entropy 
might not be as useful as claimed by its proponents.
In passing we also make some critical remarks about the implementation of these 
methods. 
\end{abstract}
pacs{87.90.+y; 05.45.Tp; 87.19.Nn}
\end{titlepage} 
                  
\section{Introduction}

Since Shannon's classical works, information theoretic concepts have found 
many applications in practically all fields of science. In particular, tools 
derived from information theory have been used to characterize the degree 
of randomness of time sequences, and to quantify the difference between 
two probability distributions. Indeed there are a number of constructs 
which qualify as distances between two distributions. Although the 
{\it Kullback-Leibler} (K-L) relative entropy \cite{gray,guiasu} is not a distance 
in the mathematical sense (it is not symmetric), it plays a central role
as it has numerous applications and numerous physical interpretations.
Another, seemingly independent, observable measuring a dissimilarity between 
two distributions was recently introduced in 
\cite{saparin}. This ``renormalized entropy" was subsequently applied to 
various physiological time sequences, including heart beats \cite{kurths,voss} 
and electroencephalograms (EEGs) recorded in patients with epilepsy\cite{timmer}.
The relation between K-L and renormalized entropy, and their application to 
EEGs recorded in patients with epilepsy is the subject of the present communication.

Ever since the first recordings in the late '20s, the EEG is one 
of the most powerful tools in neurophysiology \cite{lopes}. An
important application of EEGs in clinical practice is the diagnosis of epilepsy.  
Characteristic abnormal patterns help to classify epilepsies, to localize 
the epileptogenic focus, and eventually to predict seizures \cite{klaus}. 
About $20\%$ of patients suffering from focal epilepsies do not improve with 
antiepileptic medication and are therefore assumed candidates for a 
surgical resection of the seizure generating area. 
Successful surgical treatment of focal epilepsies requires exact
localization of the seizure generating area and its delineation from functionally relevant areas.
Recording the patient's spontaneous habitual seizures
by means of long-term (several days), and in some cases intracranial, EEGs 
(i.e., with electrodes implanted within the skull) is currently assumed most reliable. 

Although EEG recordings are in clinical use for more than half a century, conventional
EEG analysis mostly rely on visual inspection or on linear
methods as the Fourier Transform (see e.g. \cite{dummermuth} for a
comprehensive description of Fourier analysis in EEGs).
Particularly for diagnosis of epilepsy, quantitative methods
of analysis are in need to give additional information 
(for a review of quantitative methods in EEG analysis, see e.g. 
\cite{lopes}). It is precisely in this context that the authors of 
\cite{timmer} found renormalized entropy to be much more significant than 
any of the other methods they looked at.

In the following we argue that renormalized entropy is very closely 
related to K-L entropy. Indeed, it {\it is precisely} a K-L entropy, 
although not between the two distributions one started out to compare.
Nevertheless we can relate renormalized entropy 
to the K-L entropy between these two distribution. 
Moreover, when extracting 
these measures from EEGs, we find both to be very similar.
It seems indeed from these analyses that standard K-L entropy is more 
useful than renormalized entropy.

In the next section we recall Shannon and K-L entropies, and show how 
renormalized entropy is related to K-L entropy. In section \ref{sec-eeg} 
we present applications to seizure EEG data. In this section we also 
address several technical points concerning the implementation in case 
of EEG data, and we discuss the importance of the results from a 
neurophysiological point of view. Finally in section
\ref{sec-conclusions} we draw our conclusions.

\section{Entropy measures}
\label{sec-entropies}

We consider a discrete random variable having $n$ possible outcomes
$x_k~~(k=1,\ldots,n)$ with respective probabilities $p_k$, satisfying
$p_k \ge 0$ and $\sum_{k=1}^{n}~p_k~=~1$.
The Shannon entropy of $p$ is defined as 
\cite{shannon}
\be
   H[p] =  - \sum_k\; p_k\, \log p_k\;.             \label{eq:shannon}
\ee
In the following we shall take $k$ as a frequency index and $p_k$
as a normalized spectral density, 
\be
   p_k = {S(\omega_k) \over \sum_k S(\omega_k)} \;.
\ee
Moreover, the spectrum will be estimated from gliding windows over 
a scalar (`univariate') time sequence $x_n$,
\be
   S(\omega_k) = S_t(\omega_k) = \left[ |X_t(\omega_k)|^2\right]_{smooth}\;,
\ee
where $X_t(\omega_k)$ is the discrete Fourier transform of $x_n$ 
taken over a window of length $T$ centered at time $t$ (see Sec. 3 
for details), and the bracket $[\cdot]_{smooth}$ indicates a local 
averaging over nearby frequencies. We should stress, however, that 
all results of the present section apply to any probability 
distribution.

Shannon entropy is equal to $0$ in the case of delta distributions, and positive 
else. It can be interpreted as the average amount of code length (measured 
in bits, if the logarithm in eq.(\ref{eq:shannon}) is taken with base 2)
needed to encode a randomly chosen value of $k$ (randomly with respect 
to $p$). The essential point here is that the minimal (average) code
length is obtained by codes which are optimal for a specific
probability distribution -- see e.g. the Morse code which uses shorter
codes for the more frequent letters. 

Let us now suppose we have two different probability distributions $p=\{p_k\}$ 
and $q=\{q_k\}$. We can then define the K-L (relative) entropy as \cite{gray,guiasu}
\be
   K(p|q) = \sum_k \;p_k \; \log \frac{p_k}{q_k} \;.    \label{eq:kullback}
\ee
It is also positive and vanishes only if $p_k \equiv q_k$, thus
measuring the degree of similarity between both probability distributions.
Notice however, that it is in general not symmetric, $K(p|q) \neq K(q|p)$, 
therefore it is not a distance in the usual mathematical sense.
Its most important interpretation is the following: Assume that $p$ is 
the correct distribution, but the encoding is made using a code which would 
have been optimal (i.e., would have produced the shortest average code 
length) if the distribution were $q$ instead. Then, $K(p|q)$ measures
the average excess of the code length (again measured in bits, if the logarithm
is base 2) over the shortest code (which would have been based on $p$). 
But there are also several different interpretations in different 
contexts. For instance, mutual information \cite{gray} can be considered as 
K-L entropy with $p$ the true joint distribution and $q$ the 
product of the marginal distributions.
Also, Boltzmann's H theorem is most easily derived using K-L entropies
\cite{guiasu}.

A supposedly different and independent distance measure between two distributions
was introduced in \cite{saparin}. These authors called $q$ the `reference 
distribution'. They defined a `renormalized' reference distribution 
$\tilde{q}$ as
\be
   \tilde{q}_k = C \cdot [q_k]^\beta    \label{eq:refspectrum}
\ee
where $C$ and $\beta$ are uniquely fixed by demanding
\be
   \sum_k \; \tilde{q}_k \; \log q_k = \sum_k \; p_k \; \log q_k   \label{eq:renormalized} 
\ee
and 
\be
   \sum_k \; \tilde{q}_k = 1 \;.         \label{eq:energy}
\ee
Then they define `renormalized entropy' as 
\be
    \Delta H = H[p] - H[\tilde{q}] \;               \label{eq:renentropy}
\ee
and show that it is negative definite, except when $p\equiv q$. When applying it 
to time resolved spectra of several physiological time series, it is claimed 
in \cite{saparin,kurths,voss,timmer} that $\Delta H$ gives more significant 
results (e.g., shows more clearly the onset of an epileptic 
seizure \cite{timmer}) than any other observable studied by these authors.

\noindent
We want to show now that:

(i) the renormalized entropy is just the negative of the K-L entropy 
between $p$ and $\tilde{q}$, 
\be
   \Delta H = - K(p|\tilde{q}) \;.          \label{eq:deltah}
\ee
 
(ii) the absolute value $|\Delta H|$ is less than the K-L entropy 
between $p$ and $q$, since the difference between both is also 
a K-L entropy, 
\be
   |\Delta H| = K(p|q) - K(\tilde{q}|q) \leq K(p|q)\;.  \label{eq:d-deltah}
\ee
This strongly suggests that renormalized entropy cannot be more useful than 
the standard K-L relative entropy between the unrenormalized distributions.

To prove our claims, we notice that we can rewrite eq.(\ref{eq:renormalized}), 
using eqs.(\ref{eq:refspectrum}) and (\ref{eq:energy}), as 
\be
   \sum_k \; \tilde{q}_k \; \log \tilde{q}_k =
       \sum_k \; p_k \; \log \tilde{q}_k \;.   \label{eq:ren1}
\ee
Therefore, 
\bea
   \Delta H & = & \sum_k \; \tilde{q}_k \; \log \tilde{q}_k 
                 -\sum_k \; p_k \; \log p_k \nonumber\\
            & = & \sum_k \; p_k \; \log \tilde{q}_k - \sum_k \; p_k \; \log p_k  
                = - \sum_k \; p_k \; \log \frac{p_k}{\tilde{q}_k} \;,
\eea
which proves our first claim. Furthermore, we can write 
\bea
   \Delta H + K(p|q) & = & \sum_k \; p_k \; \log \tilde{q}_k 
                          -\sum_k \; p_k \; \log q_k \nonumber\\ 
              & = & \sum_k \; \tilde{q}_k \; \log \tilde{q}_k
                   -\sum_k \; \tilde{q}_k \; \log q_k
             =  \sum_k \tilde{q}_k \log \frac{\tilde{q}_k}{{q}_k} \;,
\eea
which proves the second claim.

\section{Application to EEG data}
\label{sec-eeg}

\subsection{Details of the data}

We will illustrate the result of the previous section by
re-analyzing some of the same data used in \cite{timmer}. 
The data correspond to an intracranial multichannel EEG recording of a patient
with mesial temporal lobe epilepsy; it was sampled with 173 Hz and
band pass filtered in the range $0.53-85$ Hz.  
In Fig.~\ref{fig:data} we show EEG time sequences (500000 data points, approx. 48 min.
of continuous recording) from three different
recording sites prior to, during, and after an epileptic seizure. 
Seizure starts at about point 270000 (minute $26$) and lasts for 2 minutes. 
The recording sites are located nearest to the epileptogenic focus
(upper trace; channel abbreviation: TBAR), adjacent to the focus 
(middle trace; channel abbreviation: TR), and on the non-affected brain hemisphere 
(lower trace, channel abbreviation: TBAL)
To better visualize the dynamics, insets drawn on top of each signal
show typical EEG sequences of 10 sec duration during the pre-seizure (left),
seizure (middle), and the post-seizure stage (right).

\subsection{Power spectrum}

For a finite data set $x_n$ sampled at discrete times $t_n=n\Delta t,\; n=1,\ldots,N,
\; T=N\Delta t$, we denote by $X(\omega_k)$ its discrete Fourier
transform at $\omega_k = 2\pi k/T$, with $k=1,\ldots,N$. We estimate the power spectrum as
\be
   S(\omega_k) = C\;\sum_{n=-b}^b w(n) \cdot |X(\omega_{k+n})|^2
                                        \label{eq:discretepower}
\ee
where $w(n)$ is a smoothing function of window size $B=2b+1$,
and $C$ is a normalization factor. 
As in ref. \cite{timmer}, a Bartlett-Priestley smoothing function was used
     
\begin{equation}
w(n) \propto \left\{ \begin{array}{ll}

[1-(n/b)^2] & |n| \leq b \\
0 & |n| > b \ \ \ .

\end{array} \right.
\label{eq:bartlett}
\end{equation}

As in \cite{timmer} and for comparison purposes, we subdivide the data in 
(half overlapping) epochs of $T \simeq 24$ s ($N=4096$ data points), and choose the 
window size of the Bartlett-Priestley function as $B = 33$.
This window length corresponds to a frequency resolution of 0.042 Hz. 
In the following we consider 
the spectrum in the region $\omega < 30$ Hz since 
no interesting activity occurs outside this band \cite{gotman}.
Moreover, since we are not interested in the 
absolute power, the normalization factor $C$ is adjusted such that the
sum over all frequencies below 30 Hz gives unity.

\subsection{Shannon entropy}

Parts (a) - (c) of Figs.~\ref{fig:shannon} -~\ref{fig:kl2} show the EEG signals recorded at
the three sites, contour plots of the corresponding normalized power spectra and
time dependent estimates of the Shannon entropy $H$. 
Prior to the seizure, power spectra exhibit an almost stable but
spread frequency composition which is reflected in high values of $H$. 

When the seizure starts, the spectra in Figs.~\ref{fig:shannon} and~\ref{fig:kl1}
are dominated by a single frequency component ($\sim 7$ Hz). 
This is reflected in Fig. 2 by an abrupt decrease of $H$ by about 20\%. 
Actually, the decrease is even more pronounced 
for smaller time windows, since the period of strong coherence is much 
shorter than 24 sec. 
As the seizure evolves, the dominant frequency decreases rapidly.
This dynamics is characteristic of seizures originating from the mesial temporal lobe
(see e.g. \cite{gabor}) but it is not the only possible one \cite{spencer}. 
The rise of $H$ in both Figs. 2 and 3 immediately before the 
final drop can partially be attributed to this fast change of dynamics.
The estimated entropy is high during this phase because 
of several subsequently appearing frequencies in the same window.
The following concentration of activity at lower frequencies
finally leads to a decrease of $H$. 
To a lesser degree this is also seen in Fig.~\ref{fig:kl2}.
Within or close to the seizure generating area, $H$  
remains small throughout the entire recorded post-seizure stage.
Finally, it
slowly increases towards values that compare to those obtained during the pre-seizure stage.
Using a Shannon entropy defined from the wavelet transform, similar results were obtained in
ref. \cite{blanco} from an analysis of a scalp recorded seizure.

\subsection{Kullback-Leibler entropy}   

The time courses of the K-L entropy $K(p|q)$ are shown in parts (d) of 
Figs.~\ref{fig:shannon} -~\ref{fig:kl2}.
As reference segments we used the signals from the pre-seizure stage consisting of 4096 data points
and starting at $n=20480$. 
The sensitivity (i.e. increase of $K(p|q)$ during the seizure relative to the background level) 
is notably improved when compared to that of the Shannon entropy. 
Background fluctuations during the pre-seizure stage only slightly affected 
$K(p|q)$ since pre-seizure power spectra from different windows are almost similar. 
Also, $K(p|q)$ proved nearly independent on the choice of the reference segment, 
as long as it was chosen from the pre-seizure stage.

As with the Shannon entropy we see in Figs.~\ref{fig:shannon} and~\ref{fig:kl1} 
a marked change at seizure onset due to a concentration of spectral power 
at frequencies $\sim 7$ Hz. 
$K(p|q)$ clearly detects this difference. It also detects 
the spectral difference when lower frequencies dominate in the post-seizure stage. 
But again the rapid frequency change after seizure onset is hard
to distinguish from a broad band spectrum due to our somewhat large
window size $T$.

The last two parts of Figs.~\ref{fig:shannon} -~\ref{fig:kl2} show time courses of the
K-L entropy and the renormalized entropy calculated using a
reference segment with lowest Shannon entropy
as was done by the authors of \cite{timmer}.
For Figs.~\ref{fig:shannon} and~\ref{fig:kl1} this was after the seizure
(4096 data points starting at n=335872 and n=315392, resp.), while it was
during the seizure for data shown in Fig.~\ref{fig:kl2} (4096 data points starting at n=284672).

\begin{figure}
\begin{center}
\psfig{file=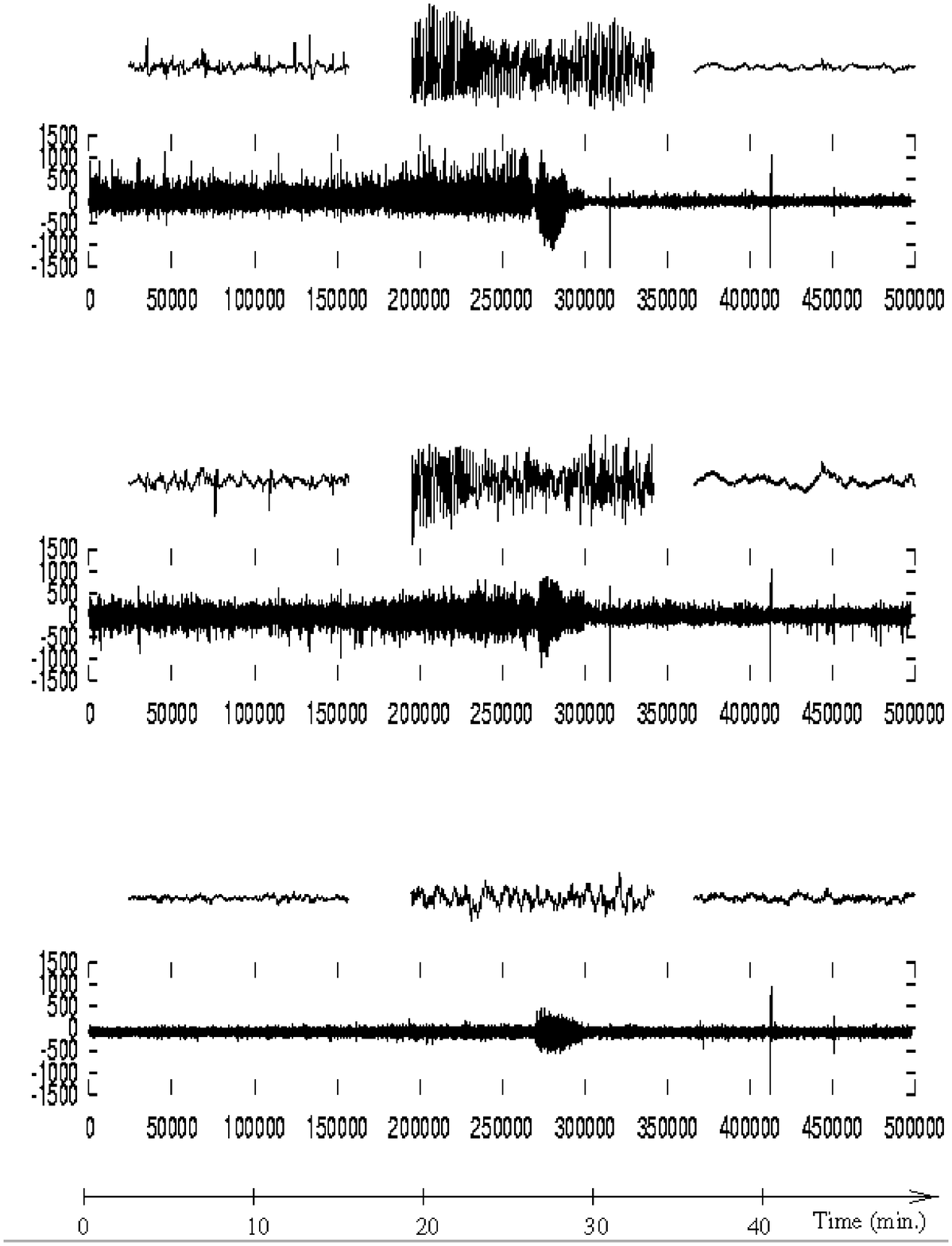,height=16cm,width=13.5cm,angle=0}
\end{center}
\caption{Intracranial EEG recordings [$\mu$V] prior, during, and after an epileptic seizure
of right mesial temporal origin. Recordings were taken from within
(electrode TBAR, upper plot) and adjacent to (electrode TR, middle plot)
the seizure generating area as well as from the non-affected brain
hemisphere (electrode TBAL, lower plot). See text
for further details. The vertical lines at about 316000, 415000 and
451000 are due to artifacts in the recording. The data corresponding to 
these artifacts were not considered for further analysis.}
\label{fig:data}
\end{figure}

\begin{figure}
\begin{center}
\epsfig{file=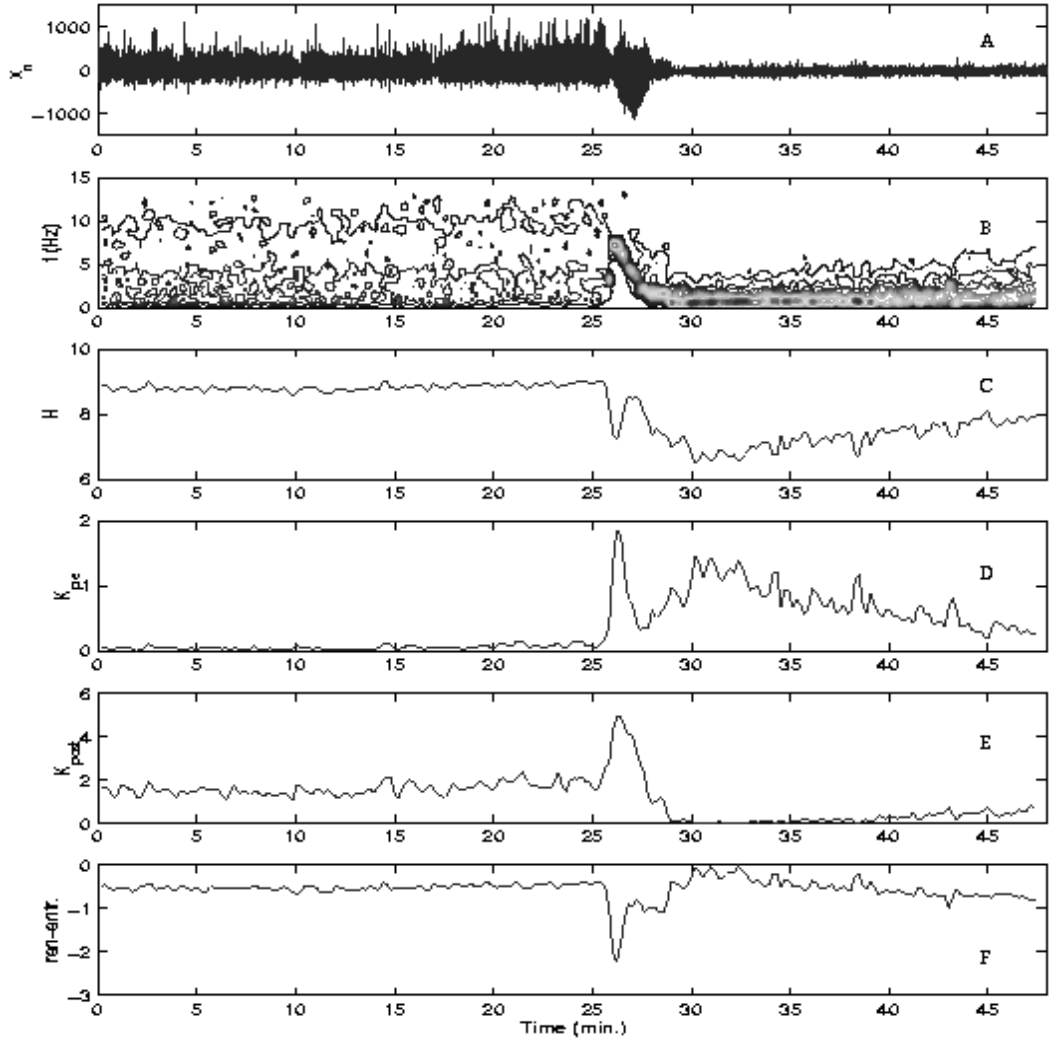,height=14cm,width=14cm,angle=0}
\end{center}
\caption{(a) EEG recording from electrode contact TBAR, (b) its corresponding power spectrum,
(c) Shannon entropy, Kullback-Leibler entropy taking a pre-seizure (d) and a post-seizure
(e) reference window, and (f) renormalized entropy (post-seizure reference window).}
\label{fig:shannon}
\end{figure}

\begin{figure}
\begin{center}
\epsfig{file=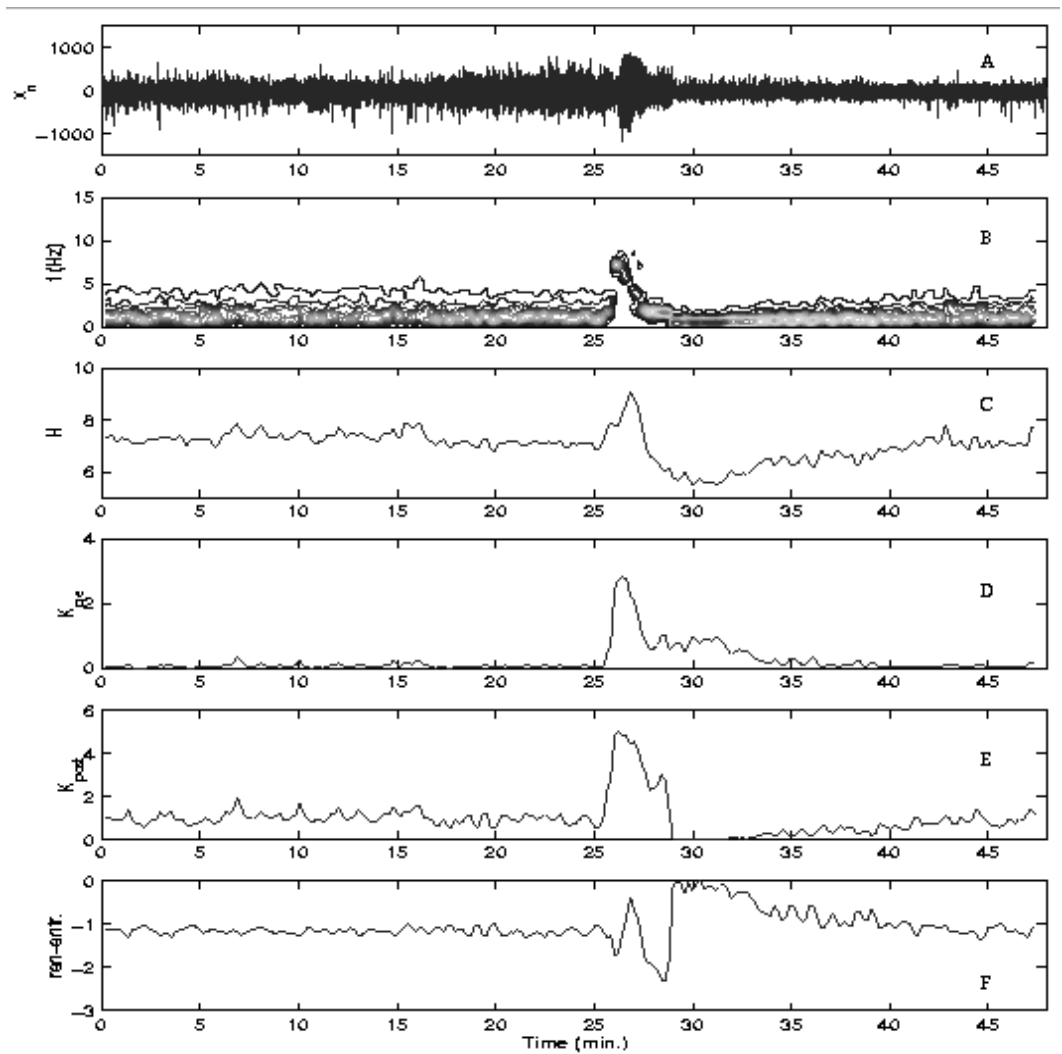,height=14cm,width=14cm,angle=0}
\end{center}
\caption{Same as Fig.~\ref{fig:shannon} but for the TR electrode.}
\label{fig:kl1}
\end{figure}

\begin{figure}
\begin{center}
\epsfig{file=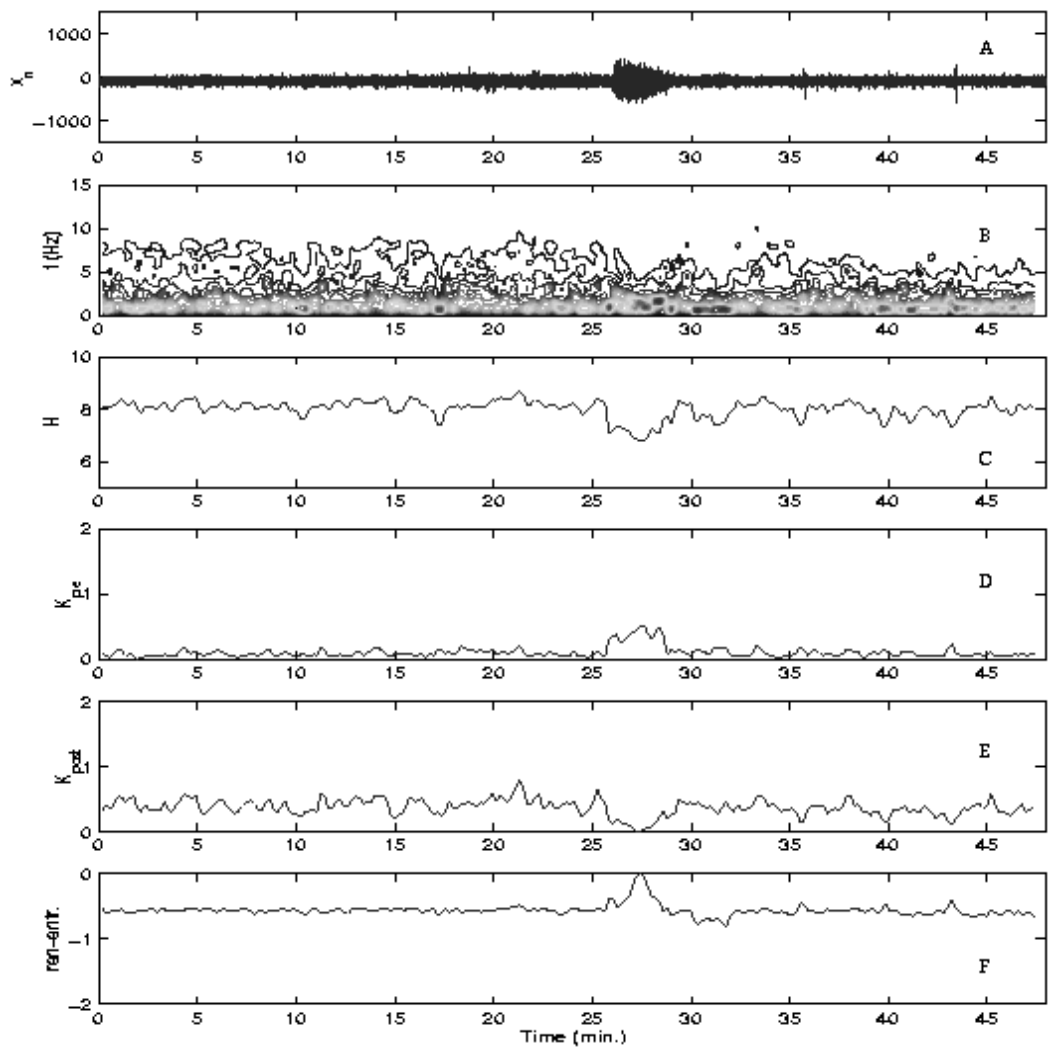,height=14cm,width=14cm,angle=0}
\end{center}
\caption{Same as Fig.~\ref{fig:shannon} but for the TBAL electrode.}
\label{fig:kl2}
\end{figure}

Here K-L and renormalized entropies give similar results. 
This illustrates the similarity between 
renormalized and K-L entropies as already pointed out in section \ref{sec-entropies}.
Differences with results in \cite{timmer} can be attributed partly 
to differences in the exact choice of the reference segment.
We see that peak values of $K(p|q)$ 
are larger than those based on calculations using a pre-seizure reference window. 
However, the relative increases over pre-seizure values are much less pronounced. 
Therefore, we consider post-seizure 
reference segments as not very useful for seizure detection. 
Moreover, post-seizure reference segments obviously can not be used in real-time applications. 
In addition, a post-seizure reference segment is not very reasonable physiologically. 
Immediately after a seizure, the state of the patient and, accordingly, the EEG are 
highly abnormal. Typically the post-seizure EEG exhibits slow fluctuations of high amplitude, 
sometimes superposed with high frequency activity (see
Fig.\ref{fig:data}). This is obviously  not an typical background EEG.
 Moreover, the post-seizure stage is often
contaminated by artifacts, some of which are not as easily recognizable as 
those shown in Fig.~\ref{fig:data}.

We therefore disagree with the procedure proposed in
ref. \cite{timmer} of automatically choosing a reference as the
segment with lowest entropy for each recording channel. 
Instead, we propose to choose a reference segment recorded during a state
as ``normal" as possible, i.e. far from a seizure (we should note, however, that 
there is still a lot of controversy in neurophysiology
of what is considered to be ``far"), free of artifacts and, if possible, free of abnormal 
alterations (admittedly, this is not always possible).
Moreover, the reference segment should be exactly the same time interval for all channels.
Otherwise comparisons between different recording sites are not reliable.
Also, one might consider taking shorter time segments. This would of course enhance statistical 
fluctuations, but would allow better time resolution. 

Even then it would be difficult  
to detect the recording site showing the very first sign of the seizure which is necessary for an exact 
focus localization. We verified this for windows down to 1.5 seconds (data not shown). This is
in agreement with clinical experience which shows that the time scales relevant for this detection 
can be less than 1 sec.
Because of these problems, the suggestions of ~\cite{timmer} concerning clinical applications like 
seizure detection or localization of epileptic foci seem too optimistic.

\section{Conclusion}
\label{sec-conclusions}

The aim of the present paper was twofold. Firstly, we showed that 
``renormalized entropy",
a novel entropy measure for differences in probability distributions,  
is closely related to Kullback-Leibler entropy. We also argued that 
it is very unlikely that more information is obtained from the former 
than from the latter. 
Secondly, we checked recent claims that renormalized entropy
(and thus also K-L entropy) is very useful in applications to 
intracranial EEGs from epilepsy patients. We found some of these 
claims to be unjustified. Nevertheless, the fact remains that K-L entropy
applied to spectral distributions is a very promising tool which 
has not yet been studied much in this context. 
In fact, ``abnormal" frequency patterns corresponding to epileptic
seizures were better identified with K-L than with the Shannon
entropy.
While the present study was performed on a limited amount of
data, we suggest K-L entropy to be an interesting tool 
for a more systematic study.

Finally, we point out that the K-L entropy can also be defined from
other time-frequency distributions rather than the windowed Fourier
transform. In particular, we consider wavelets as good candidates, since 
they have optimal resolution both in the time and the frequency range
(see \cite{grossmann,chui} for theoretical background 
and \cite{wavalpha,schiff} for application to EEGs).\\ \\

\noindent
{\bf Acknowledgments:}
K.L. acknowledges support from the Deutsche Forschungsgemeinschaft.


\newpage

\newpage

\newpage

\end{document}